\documentclass[conference]{IEEEtran}
 \IEEEoverridecommandlockouts
\usepackage{cite}
\usepackage{amsmath,amssymb,amsfonts}
\usepackage{graphicx}
\usepackage{textcomp}
\usepackage{xcolor}
\usepackage{comment}
\usepackage{optidef}
\usepackage[T1]{fontenc}
\usepackage{mathtools} 
\usepackage{cuted}
\usepackage{bbm}
\usepackage{dsfont}
\usepackage{mathtools}
\usepackage{textgreek}

\DeclareMathOperator{\Bernoulli}{Bernoulli}

\usepackage[keeplastbox]{flushend}
\usepackage[linesnumbered,ruled,vlined]{algorithm2e}

\SetKwInput{KwInput}{Input}                
\SetKwInput{KwOutput}{Output}              

\begin{document}
\title{Deep Neural Network-Based Blind Multiple User Detection for Grant-free Multi-User Shared Access
\thanks{This work was supported by the Academy of Finland 6Genesis Flagship (grant no. 318927).}}

 \author{\IEEEauthorblockN{Thushan Sivalingam, Samad Ali, Nurul Huda Mahmood, Nandana Rajatheva, and Matti Latva-Aho}
 \IEEEauthorblockA{Centre for Wireless Communications, University of Oulu, Oulu, Finland}
\{thushan.sivalingam, samad.ali, nurulhuda.mahmood, nandana.rajatheva, matti.latva-aho\}@oulu.fi, 
 }
\maketitle

\begin{abstract}
Multi-user shared access (MUSA) is introduced as advanced code domain non-orthogonal complex spreading sequences to support a massive number of machine-type communications (MTC) devices. In this paper, we propose a novel deep neural network (DNN)-based multiple user detection (MUD) for grant-free MUSA systems. The DNN-based MUD model determines the structure of the sensing matrix, randomly distributed noise, and inter-device interference during the training phase of the model by several hidden nodes, neuron activation units, and a fit loss function. The thoroughly learned DNN model is capable of distinguishing the active devices of the received signal without any a priori knowledge of the device sparsity level and the channel state information. Our numerical evaluation shows that with a higher percentage of active devices, the DNN-MUD achieves a significantly increased probability of detection compared to the conventional approaches. 
 
\end{abstract}
\vspace{3mm}
\begin{IEEEkeywords}
Complex spreading, deep learning, grant-free, massive machine-type communication, multi user detection, MUSA.
\end{IEEEkeywords}

\section{Introduction}

In the development of the wireless communications industry, massive machine-type communications (mMTC) is specified as one of the three main service classes for fifth-generation (5G) wireless communication systems and beyond~\cite{BroadbandWhitePaper2020}. mMTC supports a wide range of industrial, medical, commercial, defense, and general public applications in the Internet-of-Things (IoT) domain. The fundamental character of mMTC is to establish the uplink-dominated communication with a massive amount of machine-type devices (MTDs) to the base station (BS)~\cite{9083794}. Usually, MTDs transmit short-sized packets with low transmission rates during the short period of the active state~\cite{8663999}. Therefore conventional four-step scheduling-based multiple access schemes are inefficient as they result in a high control overhead~\cite{8663999}.

To address the impediments mentioned above, grant-free non-orthogonal multiple access (NOMA) schemes have been introduced~\cite{7263349}. Grant-free schemes allow the MTDs to communicate pilot and information symbols without acquiring a transmission grant and thereby significantly reducing the signaling overhead. By superimposing multiple MTDs' signals, the NOMA scheme allows MTDs to increase the resource utilization by utilizing the same time and frequency resources, increasing resource utilization. In this scheme, the transmit signal's orthogonality is violated because of the superposition of signals from multiple MTDs~\cite{7263349}. To mitigate the inter-user interference, NOMA employs device-specific non-orthogonal sequences. In this regard, several signature-based NOMA schemes were proposed based on device-specific codebook structures, interleaving patterns, delay patterns, scrambling sequences, and spreading sequences~\cite{8486939}. Multi-user shared access (MUSA) is introduced by ZTE~\cite{7504361} as a spreading-sequence-based NOMA scheme.

Considering that each MTD transmits information without scheduling, a device identification process is required at the BS. This process is often called multi-user-detection (MUD). In mMTC network, a single MTD is not active for a long period, and only a few devices are active in a specific time frame. Therefore, the activity vector, which has the states of active and inactive MTDs, can be represented as a sparse vector. This problem can be modeled as a sparse signal recovery problem by exploring this multi-user activity vector's sparsity. Significantly, solving this problem involves two subqueries, i) how many devices are active, and ii) which are those active devices.

A number of studies based on compressed sensing (CS) have recently been proposed to exploit the sparse characteristic of device activity~\cite{8391766, 8961111, 7462187}. The study~\cite{8391766} presents a maximum a posteriori probability (MAP) based approach to jointly identify active devices and their correspond data symbols. The authors in~\cite{8961111} proposed a generalized multiple measurement vector (GMMV) CS-approximate message passing (AMP) algorithm for joint active user detection (AUD) and channel estimation (CE) by considering the sporadic traffic nature and the virtual angular domain sparsity of massive multiple input multiple outputs (MIMO) channels. Similarly, in~\cite{7462187}, the authors proposed a structured CS-based low-complexity iterative algorithm to jointly detect active users and data. Furthermore, the authors in~\cite{8288402} proposed MUD without the reference signal at the BS for MUSA. First, the spreading codes used by the active users are estimated by an iterative algorithm. Then the active users are detected by blind equalization. 

However, with a high overloading ratio ($OR$) where the number of devices is higher than the number of resources, the performance of the previously mentioned~\cite{8391766, 8961111, 7462187} CS-based approaches degrade due to the increased correlation between the columns of the system matrix. Furthermore, conventional CS-based solutions strongly depend on the channel estimation quality. However, perfect channel estimation cannot be accomplished when the resources are limited in mMTC. The aforementioned iterative algorithm~\cite{8288402} also takes a considerable amount of time to converge, which increases the communication latency. Taken together, developing a practically feasible MUD scheme for grant-free MUSA for mMTC is a challenging open problem.

Machine Learning has been used in various domains in wireless communications, from automating ordinary tasks to advancing intelligent insights due to the advancement of computation power and advanced algorithms~\cite{6GFlagship_WP2}. In that perspective, few studies have attempted to explore the MUD problem by deep learning, e.g.,~\cite{8968401,9149252}. Both studies proposed MUD for a grant-free low-density signature (LDS) scheme. In~\cite{8968401} the authors proposed to use three deep learning-based parallel receivers with softmax sparsity estimation. Although extensive research has been carried out to detect active users, the softmax thresholding approach performs better for a specific number of active devices where the threshold matches and lead to error for unknown/unmatched numbers of active devices. Clearly, softmax produces output depending on the overall active and inactive states of the devices. But the problem mentioned above consists of independent devices.  Furthermore, using the same deep learning model in an ensemble way doesn't improve the result but increases the cost in actual implementation. In the study~\cite{9149252} the authors fixed the number of active devices during the training, which does not enable deep neural networks to learn the entire codebook of the network. Therefore, it induces misdetection during practical implementation. Therefore, there is a need to develop a practically feasible deep learning-based MUD for the MUSA scheme.  

In this work, we have proposed deep neural network-based MUD for a grant-free MUSA scheme. The major contributions of this study are as follows:
\begin{itemize}
    \item{We have introduced a novel deep learning-based MUD for grant-free NOMA scheme, specifically for the MUSA scheme. The proposed deep neural network architecture learns the correlation between the received signal and the device-specific spreading sequences.}
    \item{We have contributed jointly to detect the number of active devices and their identity by using sigmoid estimation.}
    \item{We have proposed a scalable heuristic algorithm to choose the spreading sequences with the required orthogonality factor among the possible set of MUSA spreading sequences to reduce the correlation between the MUSA spreading codes.}
    \item{We have evaluated the MUD performances of the proposed architecture and compared it with least squares-block orthogonal matching pursuit (LS-BOMP)~\cite{8570860} and Complex-AMP (C-AMP)~\cite{6478821} algorithms.}
\end{itemize}

The structure of our paper as follows. Section II presents the system model and problem formulation, including the concept of MUSA spreading sequences, generation, and selection of the relevant sequences. Section III describes the deep learning approach, MUD structure, and training and testing data generation. Section IV shows the simulation setup, parameters and, numerical results. Finally, Section V concludes the study. 
\subsubsection{Notation} 
Boldface uppercase, boldface lowercase, lower case letters represent matrices, vectors, and scalars, respectively, and calligraphy letters denote sets. \begin{math} \mathbb{R} \end{math} and \begin{math} \mathbb{C} \end{math} denote the space of real and complex numbers, respectively. The operations $(.)^H$ and $(.)^T$ denote conjugate transpose and transpose, respectively. $\Re(s)$ and $\Im(s)$ are the real and imaginary part of a complex number $s$, respectively. The identity matrix is denotes $\textbf{I}$. In addition, complex Gaussian distribution with zero mean, variance ${\sigma^2}$ is represented by \begin{math} \mathcal{CN} (0,{\sigma^2}) \end{math}. The Hadamard (element-wise) product operator is denoted by $\circ$. Also, the absolute value of the complex number $x$ is represented by $|x|$ and Euclidean norm of the vector $x$ is denoted by ${\parallel x \parallel}$. 

\section{System Model and Problem Formulation}
\subsection{System Model}
\begin{figure}[t]
    \center
    \includegraphics[width=\linewidth]{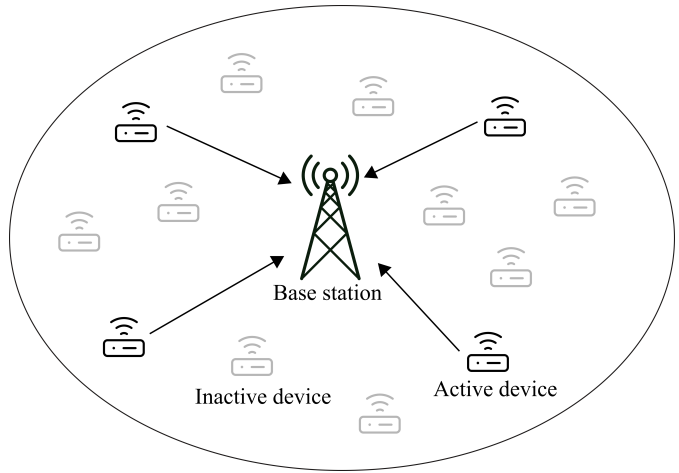}
    \caption{The illustration of mMTC grant-free NOMA uplink communication system where a fraction of devices transmits information to the BS.}
    \label{fig:System Model}
\end{figure}

Consider the uplink grant-free NOMA system of a mMTC network where a set $\mathcal{N}$ of $N$ randomly distributed MTDs are served by a BS. The BS and MTDs are each equipped with a single antenna. In this study, we focus on the overloaded scenario, which consists of higher MTDs than the available radio resources $K$ \begin{math} (K < N) \end{math}. Each MTD operates independently in this network. Furthermore, only a small subset of MTDs is active in a timeframe; the rest of the MTDs are inactive, shown in Fig.~\ref{fig:System Model}. All active MTDs communicate to the BS after spreading with the non-orthogonal complex MUSA spreading sequences. Since each active MTDs share information spontaneously without scheduling, the BS needs to classify the active MTDs, i.e., determine the number of active devices and their transmitted data. 
\subsection{Spreading Sequence-MUSA}
\begin{figure}[t]
    \center
    \includegraphics[width=\linewidth, height = 5cm]{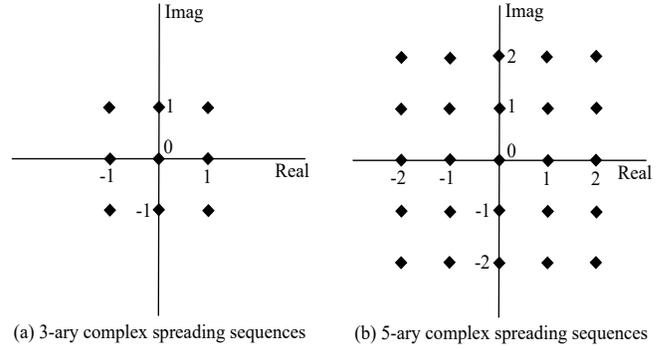}
    \caption{Elements of the MUSA complex spreading sequences~\cite{8275286}.}
    \label{fig:Spreading sequence}
\end{figure}

MUSA spreading sequences are designed to support massive MTDs in a dedicated amount of radio resources. In the MUSA scheme, each MTD information is spread with a group of complex spreading sequences. BS can generate a large set of possible MUSA codes with specific $M$-ary values and required code lengths (number of radio resources). Also, MUSA codes are designed to support high $OR$ with small code lengths. $OR$ of the MUSA spreading sequences is defined as
\begin{equation}
OR = \frac{N}{L},
\end{equation}
where $L$ is complex spreading code length (number of radio resources at the BS), above-mentioned above different $M$-ary values and code length define the space of available MUSA sequences~\cite{7504361}. The set $\{1, -1\}$ is considered as binary sequences, likewise $M = 3$ codes are generated from the element set $\{-1, 0, 1\}$ and $M = 5$ codes are generated from the element set $\{-2,-1, 0, 1, 2\}$. Then based on the element set the complex spreading elements are generated, for example $\{-1+i,-1,-1-i,i,0,-i,1+i,1,1-i\}$ is the possible complex element set for $M = 3$ codes which shown in Fig.~\ref{fig:Spreading sequence}(a). In addition, $M = 5$ codes are shown in Fig.~\ref{fig:Spreading sequence}(b). We further consider only $M = 3$ codes for our study, which is recommended by the MUSA designers~\cite{7504361}. 

After generated the complex spreading elements, we can create the space of all possible MUSA sequences by permuting them according to the required code length. Therefore, $9^L$ complex sequences can be generated from $M = 3$ codes. However, we need only $N$ number of MUSA spreading sequences. Out of all available sequences, some are highly cross-correlated sequences, which reduces the MUD performance. Rather than choosing random codes, we proposed a scalable heuristic algorithm to choose a required number of low-cross correlated sequences, summarised in Algorithm 1. 
\begin{algorithm}[ht]
  
  \KwInput{Cross-correlation threshold $\rho$}
  \KwOutput{A matrix $\tilde{\mathcal{M}}$ of MUSA sequences with low mutual cross-correlation}
  \KwData{Matirx ${\mathcal{M}}$ with all possible combination of MUSA sequences}
  \textbf{Initialization:} Candidate columns $\mathcal{C} = \mathcal{M}; \tilde{\mathcal{M}} = \emptyset $\;
\While{$ \mathcal{C} $}
   {
           $\mathbf{m}$ = randomly selected column from $\mathcal{C}$\;
           $\tilde{\mathcal{M}} = \tilde{\mathcal{M}} \cup \mathbf{m}$\;
           $\mathcal{C} = \left\{ \mathbf{c} \mid \mathbf{c} \in \mathcal{C}, \left(\frac{\mathbf{c}}{\|\mathbf{c}\|}\right)^T\left(\frac{\mathbf{m}}{\|\mathbf{m}\|}\right) \leq \rho \right\}$
          
   }

\caption{A heuristic algorithm to select low-cross correlated MUSA sequences}
\end{algorithm}
\subsection{Problem Formulation}
We define the state indicator (active or inactive) $\psi_i$ of the $i$-th MTD as
\begin{equation}
 {\psi_i} = 
\begin{cases}
1, \ $ $i$th MTD is active,$\\
0, \ $ $i$th MTD is inactive.$
\end{cases}
\end{equation}
The probability of the active state of the $i$-th MTD is $p_i$, which is independent of other devices in the cell. The received signal at the BS can be expressed as
\begin{equation}
\textbf{y} = \sum_{i=1}^{N}{\psi_i}{\textbf{s}_i}{{h}_i}{x}_i + \textbf{w},
\end{equation}
where \begin{math} {\textbf{s}_i} \in \mathbb{C}^L \end{math} is the MUSA spreading sequence vector of the $i$-th MTD, $h_i$ is the complex uplink channel coefficient from the $i$-th MTD to the BS, $x_i$ is the transmit symbol of the $i$-th MTD, and \begin{math} \textbf{w} \sim \mathcal{CN} (0,{\sigma_w^2}\textbf{I})\end{math} is the complex-Gaussian noise vector. As specified earlier, at the beginning of the transmission, BS does not have knowledge of the number of active devices and corresponding spreading sequences. Therefore BS performs MUD to identify them. In particular, all the active MTDs transmit the pilot symbol $x_{p,i}$ to the BS. Then BS detects the active devices and corresponding spreading sequences ($\textbf{s}_i$). After that, it estimates the channel coefficients ($h_i$). Finally, BS decodes the $J$ data symbols $x_{d,i}^{[1]}, \dots, x_{d,i}^{[J]}$ transmitted by the active MTDs. The transmission is over $J+1$ slots, and we assume the channel to remain the same over this duration. The pilot measurement vector \begin{math} \textbf{y}_{p} \end{math} is given by
\begin{equation}
\textbf{y}_{p} = \sum_{i=1}^{N}{\psi_i}{\textbf{s}_i}{{h}_i}{x}_{p,i} + \textbf{w}_p.
\end{equation}
Let $\boldsymbol{\phi}_i$ = $\textbf{s}_i$${x}_{p,i}$, and $\boldsymbol{\Phi}$ = [{$\boldsymbol{\phi}_1$}, \dots, {$\boldsymbol{\phi}_N$}], we then have 
\begin{equation}
    \textbf{y}_{p} = \sum_{i=1}^{N}{\boldsymbol{\phi}_i}{\psi_i}{{h}_i} + \textbf{w}_p = \Phi(\boldsymbol{\psi} \circ \textbf{h}) + \textbf{w}_p,
\label{eqyp}
\end{equation}
where $\boldsymbol{\psi}$ = $[\psi_1, \dots, \psi_N]^T$ is the activity vector, and \textbf{h} = $[{h}_1, \dots, {h}_N]^T$ is the channel vector. Let us define $\boldsymbol{\varphi} = (\boldsymbol{\psi} \circ \textbf{h}) = [\psi_1h_1, \dots, \psi_Nh_N]^T$. We can rewrite (\ref{eqyp}) as
\begin{equation}
    \textbf{y}_{p} = \boldsymbol{\Phi}\boldsymbol{\varphi} + \textbf{w}_p.
\end{equation}
We consider a fraction of MTDs ($n$ out of $N$) to be active. The accumulated sparse vector $\boldsymbol{\varphi}$ has $n$ non zero elements. Therefore, the received vector $\textbf{y}_{p}$ can be represented as a linear combination of $n$ submatrices of ${{\phi}_1 \dots {\phi}_N}$ perturbed by the noise. In MUSA sequence-based NOMA schemes, each MTD is able to choose the spreading sequences independently, which ease the resource coordination of the BS. However, BS knows all the available spreading sequences and pilot symbols. Clearly, the objective of the BS is to detect the $n$ submatrices in the received signal vector. For instance, if the second and the third codes are taken by active MTDs, then $\boldsymbol{\phi}_2$ and $\boldsymbol{\phi}_3$ are the components of $\textbf{y}_{p}$. Consequently, the MUD intricacy can be expressed as the sparse signal recovery problem: 
\begin{equation}
       \Omega = \arg\min_{\mid\Omega\mid=n} \frac{1}{2} \parallel \textbf{y}_p- \boldsymbol{\Phi}_\Omega\boldsymbol{\varphi}_\Omega \parallel_2^2.
\end{equation}
\section{Proposed Deep Learning based MUD}
\subsection{Proposed Solution Approach}
This research intends to propose the possible implementation of mMTC by enabling reliable and effective MUD under the grant-free MUSA scheme. Considering mMTC traffic, the MTD activity model is typically sporadic; therefore, only a small subgroup of the entire group of MTDs is active in a given time frame. By employing this fact, the MUD problem can be expressed as a deep learning-based supervised multi-label classification. 

The objective of the deep neural network (DNN) based MUS is to detect the corresponding labels of the $\boldsymbol{\varphi}$. Here, the number of labels and their positions will be the number of active devices and their spreading sequences. Therefore, we are not focusing on the recovery of the nonzero components. Since we train our network with sufficient training data, we don't need to estimate the channel before the detection. Therefore, we introduce a feedforward DNN based architecture to solve the tasks as mentioned above jointly. This architecture consists of several hidden nodes connecting the input received signal and the output. Here each data item from the training data ($\textbf{y}_{p}$) set is tagged with the corresponding label ($\boldsymbol{\psi}$). In order to solve the two subtasks, DNN learns the mapping function linking the input nodes and related labels by updating the hidden parameters using the backpropagation process. Clearly, It is the nonlinear mapping $g$ between $\textbf{y}_{p}$ and sparse vector elements of $\boldsymbol{\varphi}$. Consequently, we reformulate the problem (7) as 
\begin{equation}
    \Omega = g(\textbf{y}_{p};\Xi,\Delta),
\end{equation}
where $\Xi$ is the set of weights of the hidden layers, and $\Delta$ is the set of biases of the hidden layers of the neural network. The explicit intention of this DNN-MUD is to obtain $g$ characterized by $\Xi$ and $\Delta$ given $\textbf{y}_{p}$, nearest to the unknown actual mapping function $g^*$. Therefore, we model our DNN-MUD to thoroughly understand the correlation of the sensing matrix $\boldsymbol{\Phi}$. Moreover, mutual coherence is the measure of the correlation between two columns of a matrix which is defined as 
\begin{equation}
    \zeta(\boldsymbol{\Phi}) = \max_{i\neq j}|{\Phi}_i^H{\Phi}_j|.
\end{equation}
In order to learn the correlation structure of the matrix $\boldsymbol{\Phi}$ and to improve the MUD performance, we consider the restricted isometry property (RIP)~\cite{1542412} of the sensing matrix $\boldsymbol{\Phi}$. There exists a restricted isometric constant (RIC) \begin{math}\delta_s \in (0,1) \end{math} satisfying 
\begin{equation}
    (1-\delta _{s}) \Vert \boldsymbol{\varphi} \Vert_2^{2} \leq \Vert \boldsymbol{\Phi}\boldsymbol{\varphi}  \Vert_2^{2} \leq (1 + \delta _{s}) \Vert \boldsymbol{\varphi} \Vert_2^{2}.
\end{equation}
Hence, a smaller $\delta_s$ achieves better MUD performance. Therefore we consider the reduction of RIC by the design of our DNN-MUD architecture.  

\subsection{Deep Neural Network-AUD Structure}
The proposed DNN-MUD consists of dense layers, dropout layer, rectified linear unit (ReLU) activation unit, sigmoid function, and batch normalization. We use the set $\mathcal{D}$ of $D$ training data $(\Tilde{\textbf{y}}_{p}^{(1)},\dots,\Tilde{\textbf{y}}_{p}^{(D)})$ for the training phase. Here, $\Tilde{\textbf{y}}_{p}^{(d)}$ is a complex vector that cannot input directly into the DNN because our label is not in the complex form. Therefore, we divide the real and imaginary components separately and stack them as an input vector to the deep learning model as
\begin{equation}
    \Tilde{\textbf{y}}_{p}^{(d)} = [\Re(\Tilde{y}_{p,1}^{(d)}), \dots ,\Re(\Tilde{y}_{p,L}^{(d)}),\Im(\Tilde{y}_{p,1}^{(d)}), \dots ,\Im(\Tilde{y}_{p,L}^{(d)})]^T.
\end{equation}
The output vector \begin{math}\Tilde{\textbf{z}}^{(d)} \in \mathbb{R}^{\upsilon \times 1}\end{math} of the dense layer stated as
\begin{equation}
    \Tilde{\textbf{z}}^{(d)} = \textbf{W}^{in}\Tilde{\textbf{y}}_{p}^{(d)} + \textbf{b}^{in}, \ \text{for} \ d = 1, \dots, D,
\end{equation} 
where \begin{math}\textbf{W}^{in} \in \mathbb{R}^{\upsilon \times 2K} \end{math} is the initial weight, \begin{math}\textbf{b}^{in} \in \mathbb{R}^{\upsilon \times 1}\end{math} is the initial bias and $\upsilon$ is the expressing the width of the hidden nodes. After that, $D$ number of resulting vectors are assembled in the mini-batch $\textbf{B}_m$. Then, we add the batch normalization layer, where each element ${z}_j^{(d)}$ in $\textbf{B}_m$ is normalized to have zero-mean and unit-variance. The resulting vector $\Hat{\textbf{z}}^{(d)}$ of the batch normalization layer~\cite{pmlr-v37-ioffe15} is stated as  
\begin{equation}
    \Hat{z}_j^{(d)} = \gamma\left(\frac{z_j^{(d)}-\mu_{B,j}}{\sqrt{\sigma_{B,j}^2}}\right) + \eta,\ \text{for} \ j = 1, \dots, \upsilon,
\end{equation}
where $\gamma$ and $\eta$ are scaling and shifting parameters, respectively. $\mu_{B,j}$ and $\sigma_{B,j}^2$ are the batch-wise mean and variance, respectively. The function of the normalization layer is to ensure that the input distribution has a fixed mean and variance, which improves the learning of the DNN. Also, larger variances of the training data reduce the learning of the internal features from the input signal. Therefore, batch normalization controls the interpretation affected by the various wireless channels and noise. Furthermore, the ReLU activation~\cite{10.5555/3104322.3104425} function is expressed by $f(\chi) = max(\chi,0)$.

The estimation of the  $\Omega$ depends on the activation patterns of the hidden units. Furthermore, if the sensing matrix $\boldsymbol{\Phi}$ is less correlated (low mutual coherence), then the detection of the active devices would be comparatively straightforward. However, when $\boldsymbol{\Phi}$ is highly correlated, the estimation is not easy. Therefore, we introduce a dropout layer to control overfitting where some hidden nodes are randomly subsampled. Consequently, the correlation of the devices would well be learned by the network. We define the dropout vector for $t$-th layer as $\textbf{q}^{(t)}$, then $j$-th component $q_j^{(t)}$ of the vector $\textbf{q}^{(t)}$ can be expressed as 
\begin{equation}
     q_j^{(t)} = \Bernoulli(P_{dr}),
\end{equation}
where $P_{dr}$ is the dropout probability of the Bernoulli random variable. Finally, the output vector $\textbf{z}^{out}$ of the DNN is given by
\begin{equation}
    \textbf{z}^{out} = \textbf{W}^{out}\textbf{y}_{p}^{(d)} + \textbf{b}^{out},
\end{equation}
where \begin{math}\textbf{W}^{out} \in \mathbb{R}^{N \times \upsilon} \end{math}, and \begin{math}\textbf{b}^{out} \in \mathbb{R}^{N\times 1} \end{math} are the corresponding weight and bias, respectively. After that, the sigmoid function produces $N$ independent probabilities based on the previous layer. The probability of the $j$-th MTD is given by
\begin{equation}
    \hat{p_j} = \frac{1}{1+e^{-z^{out}_j}},\ \text{for} \ j = 1, \dots, N.
\end{equation}
Finally, DNN detects the active devices based on the sigmoid probability. The output estimation is given by 
\begin{equation}
    \Tilde{\Omega} = \arg\max_{\mid\Omega\mid=n} \sum_{j\in \Omega}\hat{p}_j.
\end{equation}
Here, the value of $n$ and their positions solve the previously stated two subproblems. DNN learns the sparsity by properly annotated training labels during the training phase. Therefore, DNN applies (17) to the testing data and labels to detect the active MTDs.

\subsection{Deep Neural Network-AUD Training Data Generation}
The proposed DNN-MUD is a type of supervised learning. It is a multi-class classification problem. The DNN takes the labeled data as input and learns the mapping between the data and the corresponding labels. The well-trained model is able to predict the labels for the new set of data. Therefore, a sufficient amount of training data is required for the convergence of the DNN model. We validate the adequate amount of data by the loss function of the DNN model. The proposed model can learn the parameters from the real received signal. However, we can train, validate, and test the DNN by using synthetically generated data. Furthermore, the industrial applications can validate the model using the real received signal, which is feasible since this application is for uplink communication of mMTC. 

In the training data generation process, the received signal contains the sparse input vector  $\boldsymbol{\varphi}$ and the sensing or system matrix $\boldsymbol{\Phi}$. Here, all channel properties and randomness of the environment are involved in the sparse vector, not in the sensing matrix. Furthermore, the learning part of the DNN depends on the spreading sequence matrix, which is known to the BS a priori. Therefore, synthetically generated data does not degrade the performance of the proposed DNN-MUD. 

Consequently, we generate the MUSA spreading sequences with $M = 3$ constellations for the required number of resources and devices.  Then, we generate the channel vector and random noise vector \textbf{w}. 
Finally, the training data is produced using (3) and the sparse block vector's support is used as the training label. Considering the training phase of the DNN-MUD is offline, we train the model with a different number of MTD, numbers of active MTD, and various environmental conditions. Therefore, there is no need to re-train the network frequently; it only needs to swap the trained model for different applications' requirements. 

\section{Simulation Results}
\subsection{Simulation Structure }
For the numerical evaluation, we examine the proposed DNN-MUD in the grant-free NOMA-based uplink transmission in the orthogonal frequency division multiplexing (OFDM) systems. CSI is not assumed at the BS since the BS does not know the active users a priori, and hence cannot extract the CSI from the pilot sequences. We use three $OR$ of MUSA spreading sequences to evaluate the performance of the proposed DNN based MUD scheme. In addition, we compare against the LS-BOMP~\cite{8404118} where the algorithm minimizes the ~$\ell_0$ constraint of the optimization problem and estimates the active devices and C-AMP algorithm for compression purposes. The simulation parameters are presented in Table I.
\begin{table}[ht]
\caption{Simulation Parameters}
\begin{center}
\begin{tabular}{|c|c|}
\hline
\textbf{Parameter}&{\textbf{Value}} \\
\hline
\hline
Overloading ratios ($OR$) & $200\%, 300\%, 400\%$ \\
\hline
Number of MTDs & $8, 12, 16$ \\
\hline
Active probability ($p_i$) & $1/N$\\
\hline
Number of resources (code length) & $4$ \\
\hline
MTD to BS path loss model & $128.1 + 37.6\log_{10}(r_i[Km])$  \\
\hline
Log-normal shadow fading & $8$ dB\\
\hline
Noise spectral density& $-170$ dBm/Hz  \\
\hline
 Transmission bandwidth & $1$ MHz \\
\hline
DNN sample size & $10^6$ \\
\hline
Number of dense layers & $4$ \\
\hline
Learning rate & $1\times 10^{-3}$\\
\hline
Batch size & $1 \times 10^{3}$\\
\hline
Epoch & $20$\\
\hline
Drop out, $P_{dr}$ & $0.1$\\
\hline
\end{tabular}
\label{tab1}
\end{center}
\end{table}

 For the training hyperparameters, we use the binary cross-entropy as the loss function, the Adam optimizer is used to guarantee the model's convergence, and the ReLU unit to activate the hidden layers. The proposed DNN-MUD designed, trained, validated and tested on Keras with Tensorflow backend. For the performance matrices, we use true positives (\textbf{tp}), and true negatives (\textbf{tn}), false positives (\textbf{fp}), false negatives (\textbf{fn}), precision \big($\frac{\textbf{tp}}{\textbf{tp + fp}}$\big), recall \big($\frac{\textbf{tp}}{\textbf{tp + fn}}$\big), and area under the curve/receiver operating characteristic curve to train and validate the proposed DNN-MUD model. Furthermore, we calculate the probability of detection (recall) on the testing set as $(P_D)$ = \big($\frac{\textbf{tp}}{\textbf{tp + fn}}$\big), probability of misdetection (false negative rate) as $(P_M)$ = \big($\frac{\textbf{fn}}{\textbf{tp + fn}}$\big), and positive predictive values (precision) $(PPV)$ =  \big($\frac{\textbf{tp}}{\textbf{tp + fp}}$\big) to ensure the correctness of the probability of detection. 

\subsection{Simulation Results}
\begin{figure}[t]
    \center
    \includegraphics[width=0.89\linewidth]{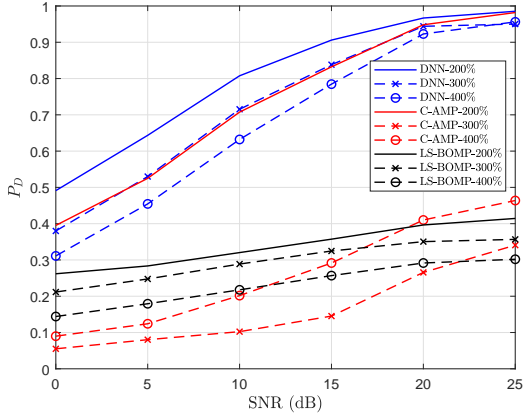}
    \caption{Probability of detection (recall) versus the SNR for three different MUD schemes with three different overloading ratios when $n=1$ of the devices are active.}
    \label{fig:DNN-MUD-1AU}
\end{figure}

First, we study the probability of detection of the proposed DNN-MUD architecture shown in Fig.~\ref{fig:DNN-MUD-1AU}; we plot the probability of detection versus signal-to-noise ratio (SNR) for $n = 1$. This result indicates that the proposed DNN-MUD performs much better than LS-BOMP and C-AMP for all three overloading ratios. Interestingly, DNN-MUD performance at 300\% and C-AMP 200\% align with each other, even though DNN performance at 200\% is higher than the two different approaches. 

\begin{figure}[t]
    \center
    \includegraphics[width=0.89\linewidth]{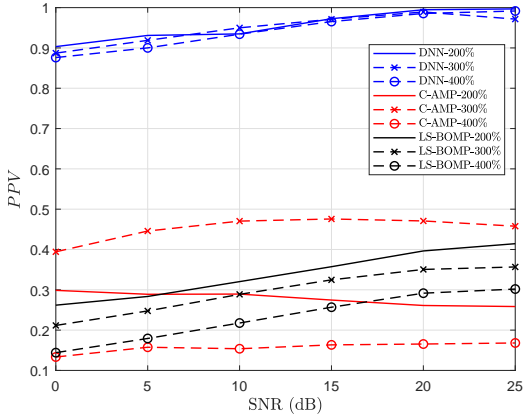}
    \caption{Positive predictive values (precision) versus the SNR for three different MUD schemes with three different overloading ratios when $n = 1$.}
    \label{fig:DNN-MUD-1AU-PPV}
\end{figure}

In Fig.~\ref{fig:DNN-MUD-1AU-PPV}, we evaluate the positive predictive value of the proposed algorithm versus SNR  for $n = 1$. This result suggests that DNN correctly detects only the relevant MTD which are active. In practice, low PPV means detecting inactive devices as active along with the active devices, which makes additional resource wastage. 

\begin{figure}[t]
    \center
    \includegraphics[width=0.89\linewidth]{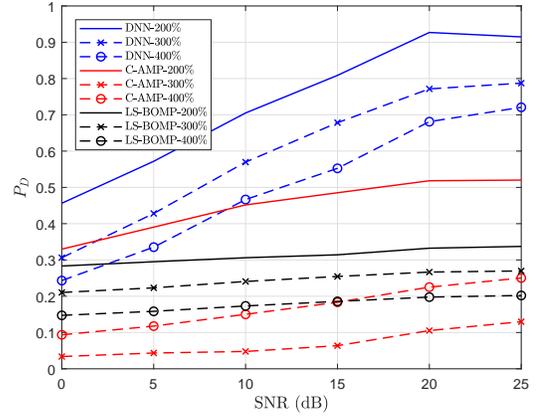}
    \caption{Probability of detection (recall) versus the SNR for three different MUD schemes with three different overloading ratios when $n = 2$ of the devices are active.}
    \label{fig:DNN-MUD-2AU}
\end{figure}

In Fig.~\ref{fig:DNN-MUD-2AU}, we examine the probability of detection versus SNR for all three different $OR$ ratios. It is a similar plot to Fig.~\ref{fig:DNN-MUD-1AU} with the difference that the number of active devices $n = 2$. We note that DNN-MUD performs better than LS-BOMP and C-AMP for all $OR$s. Specifically for 400\% $OR$, DNN-MUD exceeds the other two methods by a large margin.  

\begin{figure}[t]
    \center
    \includegraphics[width=0.89\linewidth]{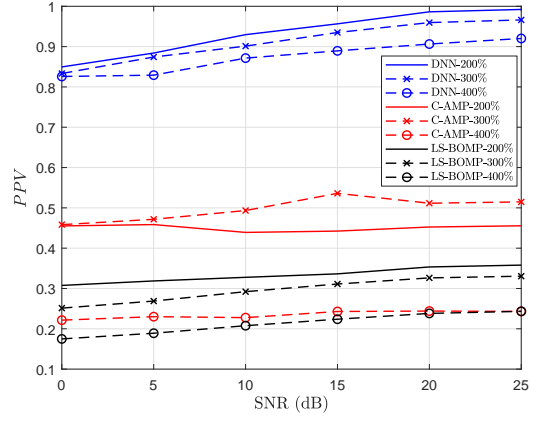}
    \caption{Positive predictive values (precision) versus the SNR for three different MUD schemes with three different overloading ratios when $n = 2$.}
    \label{fig:DNN-MUD-2AU-PPV}
\end{figure}

In Fig.~\ref{fig:DNN-MUD-2AU-PPV}, shows the positive predictive value of the proposed DNN-MUD versus SNR for $n = 2$. Taken together Fig.~\ref{fig:DNN-MUD-2AU}, we can calculate the F-measure (harmonic mean) of this problem. $F1$-score of the detection problem calculated as 
$(\frac{PPV \times P_D}{PPV + P_D})$, which means better performance of probability of detection and positive predictive value lead to the overall performance and efficiency of the system. In summary, these numerical results describe DNN-MUD outperforms all other schemes. 

\begin{figure}[t]
    \center
    \includegraphics[width=0.89\linewidth]{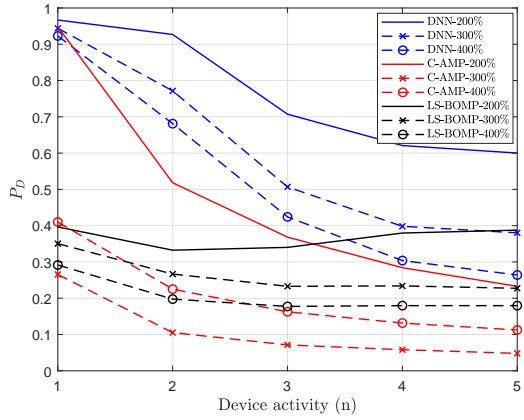}
    \caption{Probability of detection (recall) versus number of active devices for three different MUD schemes with three overloading ratios at SNR = $20$ dB.}
    \label{fig:DNN-MUD-AllAU}
\end{figure}

In Fig.~\ref{fig:DNN-MUD-AllAU}, we investigate the probability of detection versus the number of active MTDs at SNR = $20$ dB. We can see that the likelihood of detecting LS-BOMP is around $35\%$ and C-AMP is around $38\%$ for 200\% $OR$ spreading sequences when $n = 3$, while DNN-MUD achieves around 70\% of the probability of detection. Mutual coherence of the system increases sharply with the number of active MTDs. However, DNN-MUD can handle this issue by learning the sensing matrix, while LS-BOMP and C-AMP cannot. 

\begin{figure}[t]
    \center
    \includegraphics[width=0.89\linewidth]{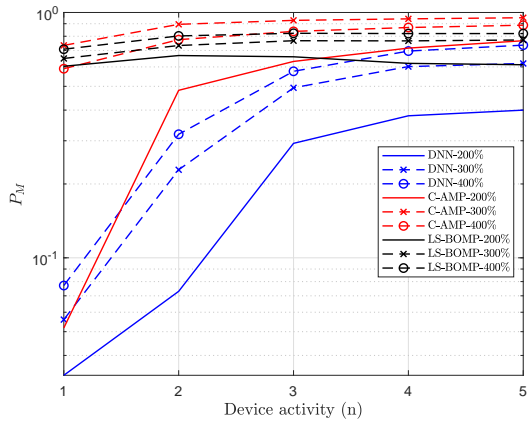}
    \caption{Probability of misdetection (false negative rate) versus number of active devices for three different MUD schemes with three overloading ratios at SNR = $20$ dB.}
    \label{fig:DNN-MUD-Mis}
\end{figure}

We examine the probability of misdetection (false negative rate) of all algorithms versus a number of active devices at SNR = $20$ dB. We plot in Fig.~\ref{fig:DNN-MUD-Mis} for three different $OR$s. Overall, DNN-MUD functions strongly than the other two approaches for three different $OR$s. Furthermore, misdetection of DNN-MUD scheme is $ P_M = 0.033$ when $n = 1$. This means $33$ misdetection in $10^3$ samples.
\section{Conclusion}
In this paper, we have introduced a novel DNN-MUD with no CSI at the BS for the grant-free MUSA in mMTC communication with sporadically transmitted short data packets and low transmission rates. To the best of our knowledge, this is the first work that presented the blind DNN-MUD for grant-free MUSA systems. First, we modeled the DNN with several hidden nodes, activation units, and a proper dropout function. Second, we used the binary cross-entropy loss function to train the DNN with the intention is to learn the structure of the sensing matrix and randomly distributed perturbations. We have provided a thorough analysis to validate the support identification performances of the proposed model. Simulation results have shown that the proposed method performs much better than the state-of-the-art conventional LS-BOMP and C-AMP methods. We have shown that DNN-MUD is a versatile design for a grant-free NOMA active device detection which can significantly improve the performance of mMTC systems. Proposing a proper deep learning model for the mobility scenarios will be an interesting problem. 
\bibliographystyle{IEEEtran}
\bibliography{main.bib}

\end{document}